\documentclass[twocolumn,amsmath,amssymb,prb]{revtex4}

\usepackage{graphicx}
\usepackage{comment}

\begin{document}
\title{Quantum oscillations in Kondo Insulator SmB$_6$
}
\author{G. Li$^1$, Z. Xiang$^{1,2}$, F.  Yu$^1$, T. Asaba$^1$,  B. Lawson$^1$, P. Cai$^{1,3}$, C. Tinsman$^1$, A. Berkley$^1$, S. Wolgast$^1$,Y. S. Eo$^1$,  Dae-Jeong Kim$^4$, C. Kurdak$^1$, J. W. Allen$^1$, K. Sun$^1$, X. H. Chen$^2$, Y. Y. Wang$^3$, Z. Fisk$^4$, Lu Li$^1$}
\affiliation{
$^1$Department of Physics, University of Michigan, Ann Arbor, MI  48109\\
$^2$Hefei National Laboratory for Physical Science at Microscale and Department of Physics, University of Science and Technology of China, Hefei Anhui 230026, China\\
$^3$Department of Physics, Tsinghua University, Beijing, China \\
$^4$Department of Physics and Astronomy, University of California at Irvine, Irvine, CA 92697, USA}

\date{\today}
\begin{abstract}
 In Kondo insulator samarium hexaboride SmB$_6$, strong correlation and band hybridization lead to an insulating gap and a diverging resistance at low temperature. The resistance divergence ends at about 5 Kelvin, a behavior recently demonstrated to arise from the surface conductance. However, questions remain whether and where a topological surface state exists. Quantum oscillations have not been observed to map the Fermi surface. We solve the problem by resolving the Landau Level quantization and Fermi surface topology using torque magnetometry. The observed Fermi surface suggests a two dimensional surface state on the (101) plane. Furthermore, the tracking of the Landau Levels in the infinite magnetic field limit points to -1/2, which indicates a 2D Dirac electronic state. 
  \end{abstract}


\maketitle                   

The recent development of topological insulators is a triumph of single electron band theory ~\cite{K1, SCZ1, MoorePRB2007, Hasan_BiSb, Shen_BiSe, Mol07,HK_RMP10,QZ_RMP11}. It is interesting to understand whether similar exotic states of matter can arise once strong electronic interaction comes into play. Kondo insulators, a strongly-correlated heavy-fermion system, offer a good playground for the exploration of this question. In a Kondo insulator~\cite{SmB6,RMPVarma}, the hybridization between itinerant electrons and localized orbitals opens a gap and makes the material insulating. Once the sample temperature is cold enough, the electronic structure in the strongly correlated system can be mapped to a rather simple electronic state that resembles a normal topological insulator~\cite{PRLcoleman}. As a result, in the ground state of the Kondo insulator there exists a bulk insulating state and a conductive surface state. In samarium hexaboride (SmB$_6$), the existence of the surface state has been suggested by recent experimental observations of the surface conductance as well as a map of the hybridization gap~\cite{Kurdak, JingXia, PRXPaglione}. However, a direct observation of the Fermi surface has not yet been achieved by transport measurements in Kondo insulators. In this letter we report the observation of quantum oscillations in Kondo insulator SmB$_6$ using torque magnetometry. The observed Fermi surface is shown to be two-dimensional (2D) and arises from the crystalline (101) surface, and the Landau Level index plot shows a Berry phase contributed -1/2 factor in the infinite field limit, which indicates that this Fermi surface encloses Dirac points, a characteristic property of topological insulators.

The direct observation of quantum oscillations is an essential step in understanding the electronic state of the bulk and surfaces of Kondo insulator. Wolgast \textit{et al.} have argued strongly that the great robustness and certain other properties of the low T surface conductivity of SmB$_6$ are best understood as a consequence of having TI surface states~\cite{Kurdak}.  Nonetheless there is yet no direct evidence for this interpretation of the surface  conduction.  Such evidence should come from microscopic measurements  of the electronic structure, as has been accomplished for the weakly correlated TI materials, such as  Bi$_2$Se$_3$, Bi$_2$Te$_3$, and graphene~\cite{Ong_interference, Ando_BiSe, Analytis_BiSe_PRB, Ong_BiTe,KimGraphene,OngBTS}.  In this report we make a microscopic study of the electronic structure of SmB$_6$ by resolving the quantum oscillations using torque magnetometry (the de Haas-van Alphen effect, or dHvA effect). 

The material SmB$_6$ is a very good insulator: the resistivity of SmB$_6$ increases by 4 orders of magnitude as its temperature decreases from room temperature to 0.3 K. For a normal insulator with such a property, usually there is no Fermi surface and no quantum oscillations. Yet, our magnetization measurements shows clear quantum oscillations due to the Landau quantization of Fermi Surface. The observed Fermi surface cross section is found to be symmetrical about a certain sample crystalline axis and is consistent with a two-dimensional Fermi surface due to the surface state of the (101) plane. The effective mass is found be as low as 0.1 $m_e$, the bare electron mass, and the mean free path is as long as 55 nm, much longer than the crystalline lattice constant.

\begin{figure}[t]
\includegraphics[width= 5 in ]{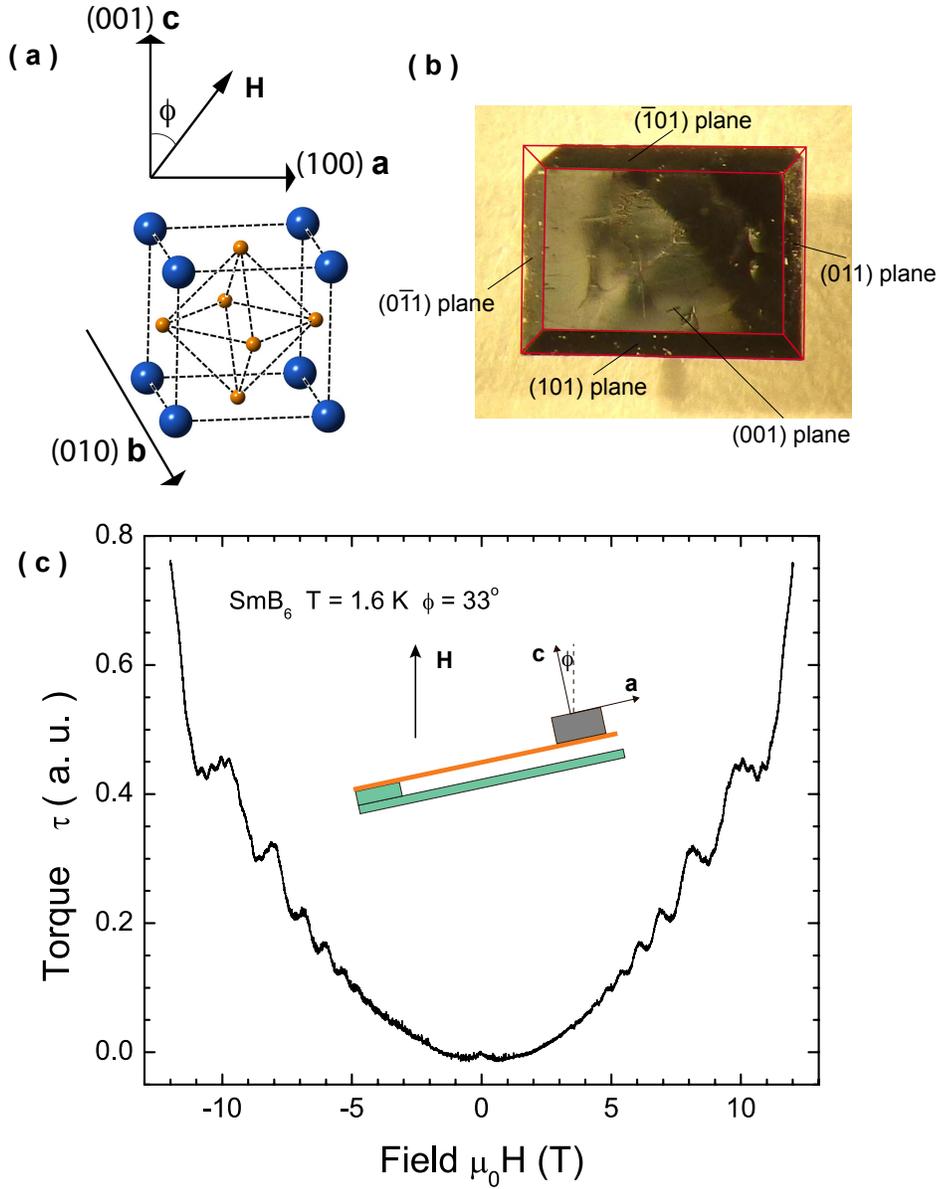}
\caption{\label{figTorque} (color online)
(a) The crystal structure of Kondo insulator SmB$_6$. The B$_6$ cluster and rare earth element Sm form a simple cubic structure. (b) The picture of the SmB$_6$ single crystal. Guide lines are drawn to label the crystalline plane of each surface. (c) The field dependence magnetic torque $\tau$ of SmB$_6$. The $\tau-H$ curve is overall quadratic, reflecting the paramagnetic state of samples. Oscillating patterns are observed in magnetic field larger than 6 T. The inset is the sketch of the measurement setup. The sample stage is rotated to tilt magnetic field $
\vec{H}$ in the crystalline $\bf \hat{a}-\hat{c}$ plane. The magnetic field is applied to the sample with a tilt angle $\phi$ relative to the crystalline $\bf\hat{c}$ axis.
}
\end{figure}

One major difference between SmB$_6$ and the conventional topological insulators is the crystal anisotropy. As shown in Fig. \ref{figTorque} (a), the crystal structure of SmB$_6$ is simple cubic. The samarium elements are separated by an inert cluster of six boron atoms.   SmB$_6$ single crystals were grown by conventional flux methods. Each sample has been etched with acid to remove the leftover flux. Fig. \ref{figTorque} (b) shows a photo of a piece of SmB$_6$ single crystal. Beside a flat (001) surface, there are four (101) planes. Later we will find one major feature of the oscillation pattern that comes from the (101) planes. 

We apply torque magnetometry to resolve the Landau Level quantization and the resulting quantum oscillation in magnetization. Since the magnetization is simply the derivative of the magnetic free energy $G$ with respect to magnetic field $H$, torque magnetometry probes the oscillations in the free energy and the density of states directly, thus making it a very sensitive probe to Fermi surface topology~\cite{Shoenberg}. Torque magnetometry measures the magnetic susceptibility anisotropy of samples \cite{NatureOscillationSebastian,LiScience08,Li13CuBi2Se3}. With the tilted magnetic field $\vec{H}$ confined to the $\hat{a}$-$\hat{c}$ plane, the torque $\tau$ of a paramagnet is shown as follows,
\begin{equation}
\tau  =  \chi_aH_aH_c-\chi_cH_cH_a  =  \Delta\chi H^2\sin\phi\cos\phi
\label{chilin}
\end{equation}
where $\phi$ is the tilt angle of $\vec{H}$ away from the crystalline $\bf\hat{c}$ axis, and $\Delta\chi = \chi_a-\chi_c$ is the magnetic susceptibility anisotropy.  Therefore, any small change of Fermi surface topology due to the Landau Level quantization is amplified by the $H^2$ term and revealed by torque magnetometry. 

In our experimental setup, an SmB$_6$ single crystal is glued to the tip of a thin brass cantilever, as shown in Fig. \ref{figTorque} (c). The magnetic torque $\tau$ is measured by tracking the capacitance change between the cantilever and a gold film underneath. An example of the torque $\tau$ vs. magnetic field $\mu_0H$  is displayed in Fig. \ref{figTorque}, taken at temperature $T$  = 1.65 K and $\phi \sim 33^{\circ}$. Overall, the $\tau-H$ curve is quadratic, reflecting the paramagnetic state and the linear $M-H$ relation in SmB$_6$. 

\begin{figure}[th]
\includegraphics[width=5 in]{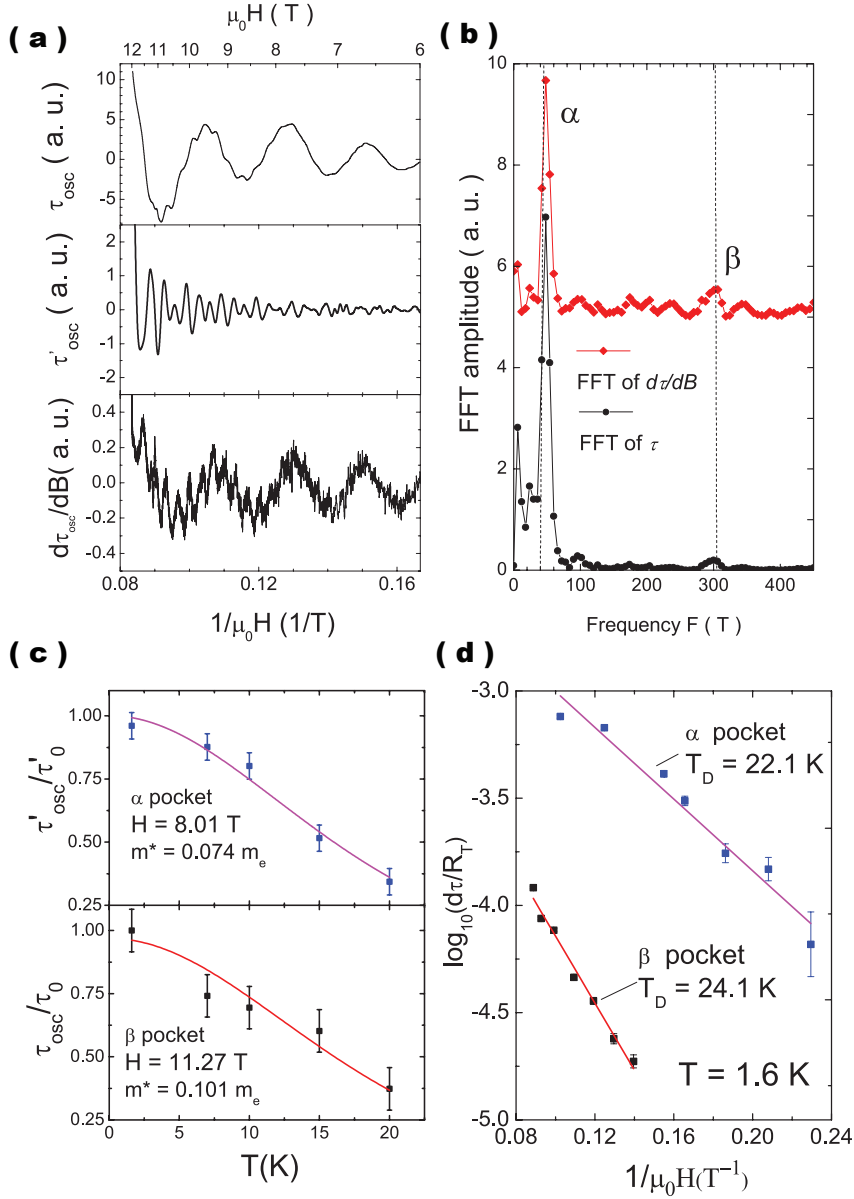}
\caption{\label{figOsc} (color online)
Quantum oscillation pattern observed by torque magnetometry. (a) The oscillatory magnetic torque s plotted as a function of 1/$\mu_0H$. $\tau_{osc}$ is taken after subtracting a polynomial background.  (b) The two main oscillation peaks are also observed in the Fast Fourier Transform (FFT) transformations of both oscillatory torque $\tau_{osc}$ as well as the derivative $\frac{d\tau}{dB}$. (c) The temperature dependence of the normalized oscillatory $\tau_{osc}$ and $\tau'_{osc}$  yield the effective mass $m = 0.074 m_e$ for the low frequency $\alpha$ pocket and $m = 0.101 m_e$ for the high frequency $\beta$ pocket. (d) At 1.6 K, the oscillating amplitude is tracked as a function of field $H$, generally known as the Dingle plot.  For the low frequency $\alpha$ pocket, the Dingle temperature $T_D$ = 22 K. For the $\beta$ pocket, $T_D$ is found to be 24 K.
}
\end{figure}

On top of the quadratic torque curve, large oscillations and small wiggles start appear as the magnetic field goes beyond 5 T. Fig. \ref{figOsc} presents the oscillation pattern. Oscillatory torque $\tau_{osc}$ is defined by subtracting a quadratic background from the torque $\tau$. Fig. \ref{figOsc}(a) shows that $\tau_{osc}$ is periodic in $1/\mu_0H$, reflecting the quantization of the Landau Levels. For metals, the oscillating frequency $F$ is determined by the cross section area $A$ of the Fermi surface~\cite{Shoenberg}, as follows,
\begin{equation}
F = \frac{\hbar}{2\pi e} A.
\label{Fs}
\end{equation}

Furthermore, there are fine wiggles in the raw torque $\tau-H$ data as well as the oscillatory $\tau_{osc}$. We further subtract the low oscillatory background away to draw attention to the high frequency oscillation. The resulting high frequency torque is named $\tau'_{osc}$, shown in the middle panel of Fig. \ref{figOsc} (a). To help determine the oscillation frequencies, we also take the derivative $\frac{d\tau_{osc}}{dH}$ and plot the result as a function of $1/\mu_0H$ in the lower panel of Fig. \ref{figOsc} (a). In the derivative plot, the slow oscillation feature lines up with that in Panel a, and the fast feature lines with that in Panel b. Further verification of the two major oscillation periods are taken by the Fast Fourier Transformation (FFT) of the oscillatory torque $\tau$, compared with that of the derivative $\frac{d\tau_{osc}}{dH}$. Two major peaks are observed in both of the FFT spectra. We name the lower frequency one as $\alpha$ and the higher frequency one $\beta$. Based on the FFT analysis and the Landau Level indexing we find the frequency $F \sim$ 48 T for Fermi pocket $\alpha$ and $F \sim$ 300 T for Fermi pocket $\beta$.

The electronic properties of these two Fermi surfaces are revealed by tracking the temperature dependence of the oscillatory torque $\tau_{osc}$ and $\tau'_{osc}$. Fig. \ref{figOsc} (c) displays the temperature dependence of the oscillating amplitude for both Fermi surfaces taken at a selected field $H$. Even though SmB$_6$ is generally known as a Kondo insulator, the temperature dependence of $\tau_{osc}$ tracks very much like that of normal metal. In metals, the oscillating magnetic torque is well described by the Lifshitz-Kosevich  formula~\cite{Shoenberg}.  The temperature $T$ and field $H$ dependences of the oscillation amplitude are determined by the product of the thermal damping factor $R_T$ and the Dingle damping factor $R_D$, as follows,
\begin{align}
R_T &= \alpha Tm^*/B\sinh(\alpha Tm^*/B) \notag \\
R_D & = \exp(-\alpha T_Dm^*/B)
\label{LK}
\end{align}
where the effective mass $m = m^*m_e$ and the Dingle temperature $T_D = \hbar/2\pi k_B \tau_S$. $\tau_S$ is the scattering rate, $m_e$ is the bare electron mass, $B = \mu_0H$ is the magnetic flux density, and  $\alpha = 2\pi^2 k_B m_e/e \hbar \sim $14.69 T/K \cite{Shoenberg}.

As shown in the $\tau_{osc} - T$ and $\tau'_{osc} - T$ plots in Fig. \ref{figOsc} (c), fitting the temperature dependence of the oscillating amplitudes of the thermal damping factor $R_T$ yields that $m = 0.074 m_e$ for Fermi surface $\alpha$ and $m = 0.10 m_e$ for Fermi surface $\beta$. We further determine the scattering rate by analyzing the Dingle damping factor. Fig. \ref{figOsc} (d) displays the field dependence of the oscillating amplitude, normalized by the thermal damping factor $R_T$. Fitting of the curves yields that $T_D \sim 22$ K for Fermi surface $\alpha$, and $T_D \sim 24$ K for Fermi surface $\beta$. Based on the parameters such as the oscillation frequencies, the effective masses, and the Dingle temperatures, we are able to determine the electronic properties of the Fermi surfaces in SmB$_6$. A summary of our results is in Table \ref{parameter}.

The measured long mean free path $l$ supports the idea of intrinsic metallic surface states. For both of the  pockets the mean free path $l$ is quite large. $l \sim 33$ nm for pocket $\alpha$ and $l \sim 55$ nm for pocket $\beta$. The mean free path is two orders of magnitude larger than the crystal lattice constant $\sim 0.4$nm. It is hard to support such a long mean free path if the surface/bulk conductance arises from hopping between impurities. 

Such a low effective mass is quite surprising, since most of the theoretical developments for the topological Kondo insulator expects a heavy mass of the surface states ~\cite{cubicTKI}. On the other hand, the relative small mass is consistent with linear Dirac dispersion and suggests the chemical potential is close to the Dirac point. For a regular quadratic dispersion, the effective mass stays the same at different chemical potential. By contrast, a linear Dirac dispersion requires a constant Fermi velocity $v_F$ at different chemical potential. The effective mass $m = \hbar k_F / v_F$ is greatly reduced once the chemical potential is very close to the Dirac point. The observed mass in SmB$_6$ is also much smaller that that of the semi metallic hexaboride compounds such as La-doped CaB$_6$ and EuB$_6$ ~\cite{PRLSdHEuB6Aronson, PRBdHvACaB6}.

\begin{figure}[th]
\includegraphics[width=5 in]{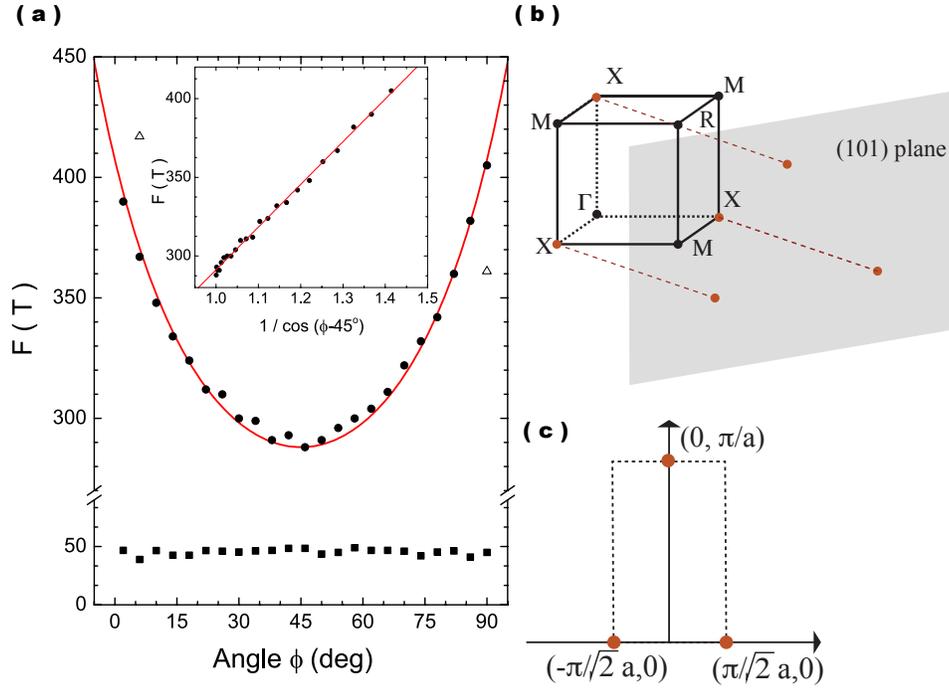}
\caption{\label{figAngle} (color online)
Angular dependence of the oscillating frequency $F$ in SmB$_6$.  (a) Oscillation frequencies of two observed Fermi surfaces $\alpha$ and $\beta$ are shown at selected $\phi$ between 0$^{\circ}$ and 90$^{\circ}$.  For both pockets, $F$  are symmetric against the $\phi = 45^{\circ}$. The red line of $\frac{F_0}{\cos (\phi-45^{\circ})}$ closely tracks the $F-\phi$ curve of the Fermi pocket $\beta$. The inset shows  $F$ of the $\beta$ pocket is proportional to $1/\cos(\phi-45^{\circ})$. Two triangle open symbols near $\phi = 0^{\circ}$ and $\phi  = 90^{\circ}$ are from the other branches from the plane (10$\bar{1}$) and ($\bar{1}$01) and they are in principle following pattern following the 2D pattern centering at $-45^{\circ}$ and $135^{\circ}$ respectively. (b) Sketch of the Brillouin Zone of SmB$_6$. At the $X$ points, the band inversion happens. The projection of these $X$ points to the (101) plane shows a possible location of the Fermi Surfaces in the (110) surface state in Panel C. (c) The projection of these X-points to the (101) plane shows the possible locations of the (101) surface state Fermi surfaces.
}
\end{figure}

The most important feature of the observed quantum oscillations is the two dimensionality (2D). For a 2D planar metallic system, the magnetic field projected to the normal axis of the plane determines the Landau Level quantizations. Thus, the oscillation frequency vs. tilt angle curve generally follows the inverse of the sinusoidal function~\cite{Ando_BiSe, Analytis_BiSe_PRB, Ong_BiTe}. We rotate the whole cantilever setup to track how the oscillation frequency $F$ changes as a function of the tilt angle $\phi$. Fig. \ref{figAngle} (a) displays the angular dependence of the oscillation frequency of both pockets. For the larger Fermi surface $\beta$, the oscillating frequency $F$ not only displays a large angular dispersion, but also closely tracks the function $1/\cos(\phi - 45^{\circ})$ (red line). To test the trend, the insert of Fig. \ref{figAngle} (a) compares $F$ vs. $1/\cos(\phi - 45^{\circ})$ for Fermi surface $\beta$ and verifies the trend with the linear fit. We also notice that the contribution from the plane (10$\bar{1}$) and ($\bar{1}$01) are also observed near $0^{\circ}$ and $90^{\circ}$.

For the small Fermi surface $\alpha$, the $F$ vs. $\phi$ also displayed a four-fold symmetry. The minimum of the oscillating period happens at $0^{\circ}$ and $90^{\circ}$. However, the scattering of the data points and the uncertainty of determining $F$ using FFT makes it challenging to make a 2D fit. 

The oscillating period of the Fermi pocket $\beta$ is symmetric to the axis (101) in the rotation plane. The observed symmetry suggests that that the oscillation does not arise from the possible remaining aluminum (Al) flux. The observed 4-fold symmetry and $1/\cos(\phi - 45^{\circ})$ can hardly be explained by Al impurity. Even if small pieces of Al flux exist in SmB$_6$ samples, they are most likely to crystallize along many random orientations. They would not likely align together to one particular crystal axis of SmB$_6$. On top of the symmetric oscillation frequency, we also find the contribution of aluminum is unlikely since the SmB$_6$ samples have all been etched to remove possible Al flux left on surfaces. Furthermore, the resistivity measurement does not show any dip at $\sim$ 1.2 K, the superconducting transition temperature of Al. These facts confirm that the observed quantum oscillation pattern is an intrinsic property of single crystalline SmB$_6$. 

The angular dependence of the oscillation frequency suggests that at least the Fermi surface $\beta$ is two-dimensional and likely to arises from the crystalline (101) plane. By contrast, most of the theoretical modeling focuses on the surface states in the (100) planes~\cite{PRLDai}. We emphasize that the surface states in the (100) plane should still be possible to observe if the future SmB$_6$ samples are cleaner. In fact, our observed Fermi pocket $\alpha$ may be from the (100) surface plane. However, this speculation does not have enough evidence due to the lack of the clear angular dependence of the measured $F$ of pocket $\alpha$. 

\begin{figure}[th]
\includegraphics[width=6.5 in]{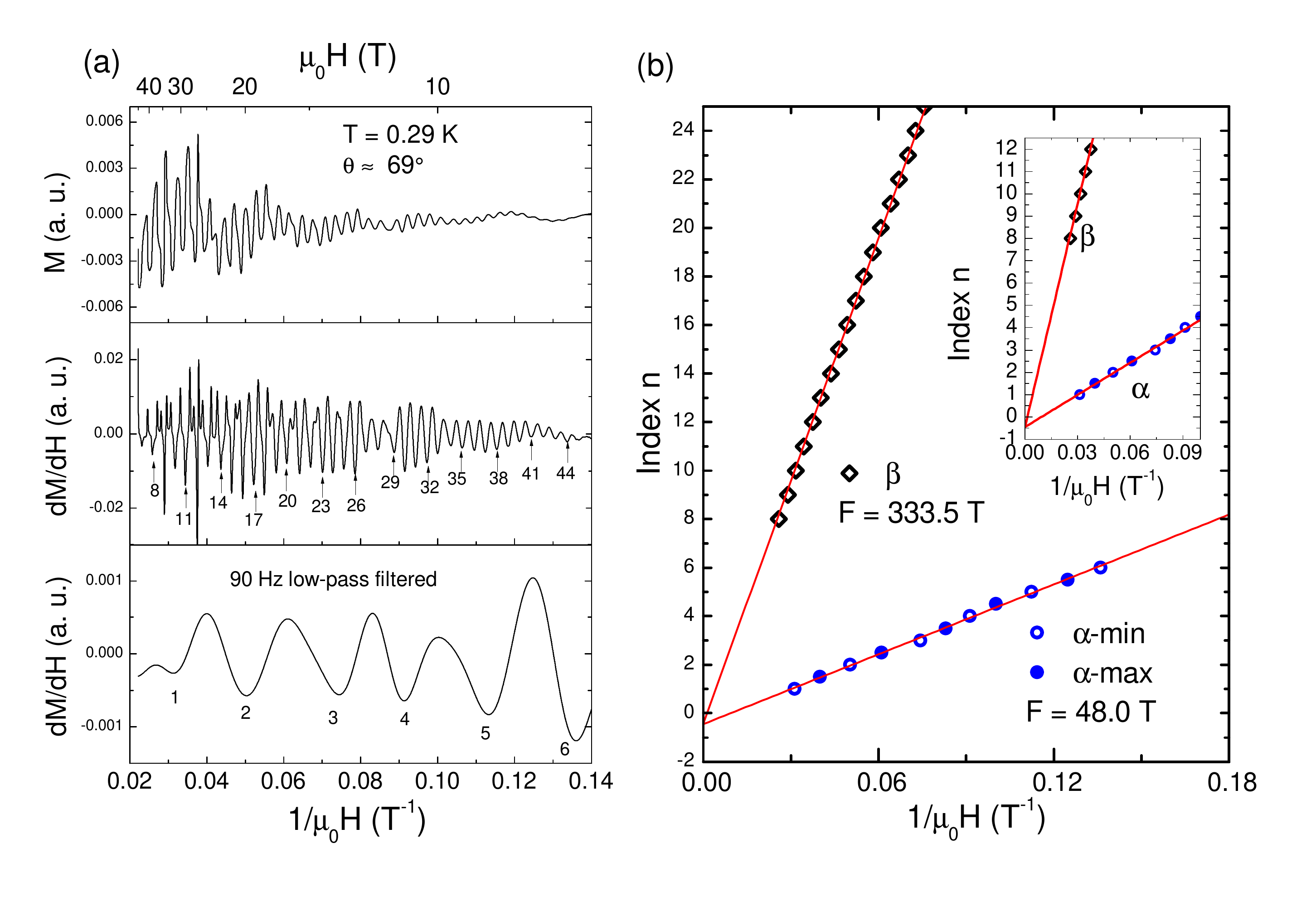}
\caption{\label{figIndex} (color online)
Landau Level index plots in SmB$_6$. (a) The magnetization $M$ of SmB$_6$ under intense magnetic field as high as $\mu_oH = 45$ T. The period structure is observed in $M - 1/\mu_oH$, and the period is consistent with the high frequency from the  $\beta$ pocket. On top of the fine structure, the oscillating pattern shows an envelope that is determined by the low frequency of the $\alpha$ pocket. The center panel shows the derivative $\frac{dM}{dH}$, in which the minimum marks the $1/B^\beta_n$ value of each Landau Level $n$ for the $\beta$ pocket.  The low frequency oscillation pattern is resolved by a low frequency filtering in the lower panel of Panel a. The Landau Levels field $1/B^\alpha_n$ are marked as the minimum of the $\frac{dM}{dH}$ curves. (b) The index plot of $1/B_n$ is plotted against the Landau Level number $n$ from Fermi pocket $\alpha$ (open blue circles) and Fermi pocket $\beta$ (open black diamond). The maximum of the $\frac{dM}{dH}$ curves are also plotted against $n + 1/2$ for pocket $\alpha$ (solid blue circles). }
\end{figure}

Finally,  we carry out the torque magnetization measurement to 45 T to track the Landau level indices for the two observed Fermi pockets.  The observed magnetization $M$ is shown in the upper panel of Fig. \ref{figIndex}(a). We take the derivative $dM/dH$, shown in the middle panel of Fig. \ref{figIndex}. Following the practices well demonstrated in the dHvA effect in bismuth~\cite{LiScience08} and proofed analytically, the Landau Level field is defined by the field value of the minimum in $dM/dH$, reflecting the Landau diamagnetic susceptibility maximizes as the chemical potential finishes filling each Landau Levels. Following the practices well demonstrated in the dHvA effect in bismuth~\cite{LiScience08}], we take the derivative dM/dH shown in the middle panel of Fig. 4. The Landau Level field, defined by the field value corresponding to the minimum in dM/dH, reflects that the Landau diamagnetic susceptibility maximizes as the chemical potential finishes filling each Landau Level.

In the raw data of $dM/dH$, the fast oscillating period is well defined, and the dips show the Landau Levels fields $B^\beta_n$ for the pocket $\beta$. The envelop of the pattern shown in the $dM/dH$ pattern is from the slow oscillation due to the pocket $\alpha$. To determine the Landau Level field $B^\alpha_n$ for the slow oscillation, we apply a lower pass FFT filter ( $F <$ 90 T ) to resolve the oscillation pattern, shown in the lower pattern of Fig. \ref{figIndex}(a). The index plots for both pockets are displayed in Fig.\ref{figIndex}(b). For the $\alpha$ pocket, we slow add the field value where the maximum happens in the plot, as the function of $n$+1/2. As expected, the index plots show straight lines. To determine the infinite field limit $\gamma$, the linear fits are extended to find the x-axis intercept. The insert of Fig. \ref{figIndex}(b) shows the same index plots in expanded scale. The intercept $\gamma$ is -1/2 for both pockets. Linear fitting of the index plots also finds $\gamma^\alpha = -0.45 \pm 0.07$ and $\gamma^\beta = -0.44 \pm 0.06$. We note that the half integer intercepts are one of the characteristic features of 2D Dirac electronic systems, as demonstrated in graphene~\cite{KimGraphene} and the surface state of topological insulators~\cite{OngBTS}. 

We notice the missing of the quantum oscillation in the magnetoresistance of SmB$_6$. We have tested the resistance of SmB$_6$ at 300 mK in field as high as 45 T, but did not resolve Landau Levels. Although we do not know exactly the reason, we speculate that it is also related to the missing of the (101) plane in most of the measurement setups. A typical four-probe resistance measurement  would be a parallel contribution of four (100) planes and four (101) planes, and the contribution from the former would easily dominate given their relative large width (for the sample shown in Fig. \ref{figTorque}, the total net width of the (100) planes is about 10 times larger than that of the (101) planes). Moreover, the magnetic torque and magnetization measurements are usually more sensitive to the Landau Level quantization, demonstrated in many material systems~\cite{NatureOscillationSebastian,LiScience08,Li13CuBi2Se3}  because the magnetization is a thermodynamic quantity directly related to the free energy and the density of states. Certain sample domain structures may affect the coherence in the resistance measurement, but not on the magnetization measurements.  

Questions remain as to where the observed Fermi surface in the (101) plane arises from.  In the (101) plane there are four high symmetry points[$(0,0)$, $(0,\pi/a)$, $(\pi/\sqrt{2}a,0)$ and $(\pi/\sqrt{2}a,\pi/a)$]. For example, as shown in Fig. \ref{figAngle} (b), projecting the bulk band $X$ points to the (101) plane leads to a pocket at (0, $\pi/a$) , and a pair of pockets at ($\pi/\sqrt{2}a$,0) and ($-\pi/\sqrt{2}a$,0). Based on symmetry, only (0,$\pi/a$) shall have a Dirac point, which is consistent with the experimental observation. However, the topological theory doesn't prohibit pairs of Dirac points to arise at low symmetry points,which offers another possible origin for the observed pockets.

In conclusion, using torque magnetometry we observed Fermi-liquid-like quantum oscillations from Kondo insulator SmB$_6$. Two small Fermi surfaces were observed with light mass $m \sim 0. 07 m_e - 0.10 m_e$ and the $k_Fl$ factor as big as 55. The angular dependence reveals that the dominating oscillation pattern is symmetric to the crystalline (101) axis. The Landau Level index plot reveal the existence of the Berry phase factor in the infinite $B$ limit and indicates that the surface state is topological. Our observation indicates a clean metallic surface state along the (101) cleavage plane, and this is an ideal plane to resolve the chirality and topological properties of the surface state. 

{\bf Acknowledgement} The material is based on work supported by the start up fund and the Mcubed project at the University of Michigan (low field magnetometry), by the Department of Energy under Award number DE-SC0008110 (high field torque magnetometry) ,  by the National Science Foundation DMR-0801253 and UC Irvine CORCL Grant MIIG-2011-12-8 (sample growth). Part of the work performed at the University of Michigan (C.K. group, device fabrication) was supported by NSF grant \# DMR-1006500.  The Corbino samples were fabricated at the Lurie Nanofabrication Facility (LNF), a member of the National Nanotechnology Infrastructure Network, which is supported by the National Science Foundation The high-field experiments were performed at the National High Magnetic Field Laboratory, which is supported by NSF Cooperative Agreement No. DMR-084173, by the State of Florida, and by the DOE.

\begin{table}[h]
\caption{Parameters in the two Fermi pockets in SmB$_6$. The oscillation frequency $F$ and Fermi wavevector $k_F$ are obtained from the dHvA period. The effective mass $m$, Fermi velocity $v_F$, mean free path $l$, and the mobility $\mu$ are inferred from the $T$ and $H$ dependences of the oscillation amplitude. The Landau Level index plot yields the infinite field limits $\gamma$. }
\centering
\begin{ruledtabular}
\begin{tabular}{c c c}

                      					& $\alpha$						&$\beta$ \\
\hline
$F$ ( T ) 					&  48.3 $\pm $1.8					&300.5 $\pm$ 1.3 			\\
$k_F$ ( nm$^{-1}$)				& 0.383 $\pm$ 0.007				&0.955 $\pm$ 0.002 \\
$\frac{m}{m_e}$				& 0.074 $\pm$ 0.004				& 0.101$\pm$ 0.012 \\
$v_F$ (10$^5$ m s$^{-1}$)		& 6.0 $\pm$ 0.4					& 10.9$\pm$ 1.3\\
$l$ (nm)						& 33 $\pm$ 7						& 55 $\pm$ 16 \\
$\mu$ ($\times$10$^3$cm$^2$/V s)	& 1.3 $\pm$ 0.3					& 0.86 $\pm$ 0.26 \\
$k_Fl$						& 13 $\pm$ 3						& 53 $\pm$ 15 \\
$\gamma$					& -0.45 $\pm$ 0.07				& -0.44 $\pm$ 0.06
\end{tabular}
\end{ruledtabular}
\label{parameter}
\end{table}

\newpage

%

%
\end{document}